\documentclass[12pt]{article}
\usepackage{amssymb,amsmath,authblk}
\usepackage{epsf,cite}
\usepackage{graphicx}
\usepackage{epstopdf}

\usepackage[utf8]{inputenc}
\usepackage[russian,english]{babel}
\usepackage{setspace}

\date{\vspace{-5ex}}

\begin{document}

\title{A new review of excitation functions of hadron production in $pp$ collisions in the NICA
energy range}
\author{V.~Kolesnikov\thanks{Vadim.Kolesnikov@cern.ch}~}
\author{V.~Kireyeu}
\author{V.~Lenivenko}
\author{A.~Mudrokh}
\author{K. Shtejer}
\author{D.~Zinchenko}
\affil{Joint Institute for Nuclear Research, Dubna Russia}
\author{E.~Bratkovskaya}
\affil{GSI Helmholtzzentrum f\"{u}r Schwerionenforschung GmbH, Darmstadt, Germany; Institut
 f\"{u}r Theoretische Physik, Johann Wolfgang Goethe-Universitat, Frankfurt am Main, Germany}

\maketitle



\begin{abstract}
Data on hadron multiplicities from inelastic proton-proton interactions in the energy
range of the NICA collider have been compiled. The compilation includes
recent results from the NA61/SHINE and NA49 experiments at the CERN SPS accelerator.
New pa\-ra\-me\-te\-ri\-za\-tions for excitation functions of mean multiplicities
$\left<\pi^{\pm}\right>$, $\left<K^{\pm}\right>$, $\left<K^{0}_S\right>$, $\left<\Lambda\right>$,
$\left<p\right>$, $\left<\bar{p}\right>$
are obtained in the region of collision energies $3<\sqrt{s_{NN}}<31$~GeV.  
The energy dependence of the particle yields, as well as variation of rapidity and transverse
momentum distributions are discussed. A standalone algorithm for hadron phase space generation
in $pp$ collisions is suggested and compared to model predictions using an example of the PHQMD
generator.\bigskip

The investigation has been performed at the Laboratory of High Energy Physics, JINR\\



 
%

\end{abstract}


\newpage

\section{Introduction}
\hspace{4mm} The NICA accelerator complex is under construction at JINR (Dubna). It would offer
a record luminosity (reaching $10^{27}$~cm$^{-2}$c$^{-1}$) for heavy-ion collisions in the energy
range $4<\sqrt{s_{NN}}<11$~GeV~\cite{nica}. Proton-proton collisions at NICA can be studied
in the energy range from 4 to 25 GeV. The physics program of the MultiPurpose Detector (MPD)
at the NICA collider is aimed at experimental exploration of a yet poorly known region of the
QCD phase diagram of the highest net-baryon density with an emphasis on the nature of the transition
from hadronic to quark-gluon degrees of freedom, modification of hadron properties in dense nuclear
matter, and search for the signals about the critical end point~\cite{nica_physics}.
However, the interpretation of experimental results from nucleus-nucleus interactions showing
novel phenomena has to rely on comparison to the corresponding data from elementary collisions.
For example, the excitation function of the strangeness-to-entropy ratio, which behaves differently
in heavy-ion and $pp$ collisions, may serve as an important probe in the study of the deconfinement
phase~\cite{na49_pika} or can be related to chiral symmetry restoration in the dense hadronic
matter~\cite{brat_csr}.

Microscopic models of nucleus-nucleus collisions, which utilize multiple physics phenomena in strongly
interacting matter, are useful tools for explaining experimental results and making new
predictions. Data on hadron yields from elementary inelastic collisions
are the essential input for such kind of models permitting to establish details of the evolution
of particle inclusive production from elementary to nuclear interactions.  
Experimental studies of hadron production in $pp$ collisions have been performed at many laboratories
over 60$^{th}$-80$^{th}$ of the last century. There are also several review papers, which summarize
data and discuss the excitation function of hadron yields in a range of collision energies from
several GeV up to LHC energies (see for example~\cite{rossi,gazdzicki:1996,weber}).
These old papers, however, rely on a too broad region of energies as compared to the NICA range
and do not include the recent measurements from
CERN/SPS~\cite{na49_pions,na49_kaons,na49_ppbar,na61_hminus,na61_hadrons,na61_lambda}.
Moreover, early bubble chamber applications are typically limited to small data samples
(of the order of $10^4$ events or so) and have no particle identification,
while old spectrometer measurements were done with small solid angle devices and have no sufficient
acceptance.
In this case extraction of integrated quantities (i.e. mean multiplicities), requiring
extrapolation into the unmeasured regions of the reaction phase space, suffers from sizeable
systematic uncertainties due the scarcity and incomplete phase space coverage
(see for example discussion in~\cite{na49_kaons}).
An updated data set of hadron yields, which includes precise measurements from modern detector
setups (i.e. NA49 and NA61/SHINE experiments at the CERN/SPS), can allow getting much better
parameterizations for the collision energy dependence of hadron production in the NICA energy range.

Of particular interest is strangeness production. At the lowest NICA energy
the excitation function of the production rates of strange hadrons varies
strongly due to threshold-dominated  effects. Moreover, differences between K$^{+}$
and K$^{-}$ yields related to the underlying production mechanisms 
are developed at NICA. In this context, since the relation
between charged and neutral kaon production is also sensitive to the respective
production mechanisms, it is important to examine available data on the K$^0$ production too.  

The aim of this paper is twofold. First, in order to improve the existing elementary data base
we want to collect the most complete set of experimental data of hadron yields from $pp$ collisions
in the NICA energy range, which includes results of mean multiplicities, rapidity distributions,
and transverse spectra.
Secondly, we want to undertake a systematic study of the collected experimental results
as a function of the collision energy and obtain proper parameterizations for the energy dependence
of inclusive production cross-sections, as well as investigate the evolution of the parameters
of the hadron phase space distributions (i.e. shapes of rapidity spectra and transverse momentum
distributions). Since most bulk observables relate to the non-perturbative sector of QCD, it is
one of the main goals of this work to obtain the basis for a model independent framework
for predicting hadron yields in $pp$ collisions at NICA energies. 
Thus, the results of this study can be used as an input for detector simulation and feasibility
study at NICA.


The paper is organized as follows. In Section 2 a new compilation of experimental data on mean
hadron multiplicities is presented. The excitation functions for hadron yields are discussed
in Section 3. Parameters of the rapidity spectra and transverse momentum distributions
for hadron species are discussed in Section 4.
In Section 5 we suggest a simulation approach for generation of the hadron phase space
distributions based on the obtained parameterizations for the excitation
function of particle production yields.
 A summary in Section 6 closes the paper.

\section{Experimental data on hadron yields from $pp$ collisions}

\hspace{4mm} In this section we present a collection of experimental data on hadroproduction
in inelastic $pp$ collisions, which ensures coverage from close to the strangeness production
threshold up to the maximum NICA energy.

\begin{table}[htp]
\begin{center}
\scalebox{0.9}{%
 \small\addtolength{\tabcolsep}{-3pt}
        \begin{tabular}{|c|c|c|c|}
        \hline
        Reference & 
        $\sqrt{s_{NN}}$~(GeV) &
        $\left<\pi^-\right>$ & Error (\%) \\ \hline
\cite{gazdzicki:1996,antinucci:1973}  & 2.99  &  0.2 & 10 \\ \hline
\cite{gazdzicki:1996,antinucci:1973}  & 3.50  & 0.29 & 10  \\ \hline
\cite{gazdzicki:1996,antinucci:1973}  & 4.93  & 0.63 & 10 \\ \hline
\cite{gazdzicki:1996,antinucci:1973}  & 5.03  & 0.75 & 10   \\ \hline
\cite{antinucci:1973}                 & 5.10  & 0.72 & 10  \\ \hline
\cite{antinucci:1973}                 & 5.97  & 0.98 & 10  \\ \hline
\cite{gazdzicki:1996,antinucci:1973}  & 6.12  & 1.01 & 10 \\ \hline
\cite{na61_hminus}                    & 6.27  & 1.05 & 5    \\ \hline
\cite{na61_hadrons}                   & 6.27  & 1.08 & 19    \\ \hline
\cite{antinucci:1973}                 & 6.38  & 1.08 & 10   \\ \hline
\cite{gazdzicki:1996,antinucci:1973}  & 6.84  & 1.11 & 10  \\ \hline
\cite{antinucci:1973}                 & 6.86  & 1.11 & 10   \\ \hline
\cite{antinucci:1973}                 & 7.43  & 1.21 & 10     \\ \hline
\cite{na61_hminus}                    & 7.74  & 1.31 & 5     \\ \hline
\cite{na61_hadrons}                   & 7.75  & 1.47 & 13   \\ \hline
\cite{na61_hminus}                    & 8.76  & 1.48 & 3      \\ \hline
\cite{na61_hadrons}                   & 8.76  & 1.71 & 10    \\ \hline
\cite{na61_hminus}                    & 12.32 & 1.94 & 4   \\ \hline
\cite{na61_hadrons}                   & 12.32 & 2.03 & 9     \\ \hline
\cite{gazdzicki:1996,antinucci:1973}  & 13.90 & 2.19 & 10  \\ \hline
\cite{na61_hminus}                    & 17.30 & 2.44 & 5     \\ \hline
\cite{na61_hadrons}                   & 17.30 & 2.40 & 8      \\ \hline
\cite{na49_kaons}                     & 17.30 & 2.36 & 2     \\ \hline
\cite{antinucci:1973}                 & 19.75 & 2.82 & 10    \\ \hline
\cite{antinucci:1973}                 & 22.02 & 2.98 & 10  \\ \hline
\cite{antinucci:1973,na49_kaons}      & 30.98 & 3.44 & 10  \\ \hline
 \end{tabular}\hspace{3mm}
\begin{tabular}{|c|c|c|c|}
 \hline
        Reference & 
        $\sqrt{s_{NN}}$~(GeV) &
         $\left<\pi^+\right>$ & Error (\%)\\ \hline
\cite{gazdzicki:1996,antinucci:1973}  & 2.99 & 0.48  & 10  \\ \hline
\cite{gazdzicki:1996,antinucci:1973}  & 3.50 & 0.67  & 10  \\ \hline
\cite{gazdzicki:1996,antinucci:1973}  & 4.93 & 1.22  & 10   \\ \hline
\cite{gazdzicki:1996,antinucci:1973}  & 5.03 & 1.37  & 10  \\ \hline
\cite{gazdzicki:1996,antinucci:1973}  & 6.12 & 1.6   & 10  \\ \hline
\cite{na61_hadrons}                   & 6.27 & 1.88  & 11   \\ \hline
\cite{gazdzicki:1996,antinucci:1973}  & 6.84 & 1.88  & 10   \\ \hline
\cite{na61_hadrons}                   & 7.75 & 2.08  & 10       \\ \hline
\cite{na61_hadrons}                   & 8.76 & 2.39  & 7      \\ \hline
\cite{na61_hadrons}                   & 12.32 & 2.67  & 5  \\ \hline
\cite{na61_hadrons}                   & 17.30 & 3.11  &13     \\ \hline
\cite{na49_kaons}                     & 17.30 & 3.02  & 2    \\ \hline
\cite{antinucci:1973}                 & 22.02 & 3.56  &10     \\ \hline
\cite{antinucci:1973,na49_kaons}      & 30.98 & 4.04  &10   \\ \hline
\end{tabular}
 }
     \caption{\label{table1}The compiled results on the mean
 multiplicity of charged pions
    from inelastic proton-proton interactions at different collision energies.}
\end{center}
   \end{table}

Early data on charged and neutral hadron production were obtained in 60$^{th}$-70$^{th}$ years
of 20$^{th}$ century by bubble chamber, streamer chamber, and Time-Projection Chamber (TPC) experiments.
Later on in a series of experiments at the CERN ISR accelerator (ISR standing for Intersecting Storage
Rings) the measurements were performed by counter experiments, where the particle identification
was done using Cherenkov and Time-of-Flight (TOF) techniques.
Phase space coverage in the most of early experiments required additional
extrapolation to get the yield of hadrons in the full momentum space. The details of this procedure
are described in the original papers for most of the results presented here
(see for example~\cite{gazdzicki:1996}).

Most recently, more accurate and much more detailed data on hadron production in minimum-bias $pp$
interactions have been obtained by the NA49 and NA61/SHINE Collaborations
at the CERN/SPS accelerator (SPS stands for Super Proton\ Synchrotron).
In the NA49 ex\-pe\-ri\-ment~\cite{na49} charged hadrons are identified by energy loss measurement
in a large volume TPC tracking system covering a significant fraction
of the production phase space. Ionization loss measurements from TPC are complemented by
a TOF system, which covers the midrapidity region. Invariant yields of $\pi^{\pm}$~\cite{na49_pions},
K$^{\pm}$~\cite{na49_kaons}, and (anti)protons~\cite{na49_ppbar} were measured for the transverse
momentum interval from 0 to 2~GeV/c and for the Feynman $x$ variable within $x=[0..0.85]$.
\begin{table}[htp]
 \begin{center}
 \scalebox{0.9}{%
 \small\addtolength{\tabcolsep}{-1pt}\vspace{-15mm}
        \begin{tabular}{|c|c|c|c|}
        \hline
        Reference & 
        $\sqrt{s_{NN}}$ (GeV) &
        $\left<K^-\right>$ & Error (\%) \\ \hline

\cite{gazdzicki:1996,antinucci:1973}  & 5.03  & 0.0095 &35  \\ \hline
\cite{gazdzicki:1996,antinucci:1973}  & 6.15  &  0.036 & 14   \\ \hline
\cite{na61_hadrons}                   & 6.27  &  0.024 & 26    \\ \hline
\cite{gazdzicki:1996,antinucci:1973}  & 6.84  &  0.031 & 14   \\ \hline
\cite{na61_hadrons}                   & 7.75  &  0.045 & 11  \\ \hline
\cite{gazdzicki:1996}                 & 7.86  &  0.05  & 30   \\ \hline
\cite{gazdzicki:1996,antinucci:1973}  & 8.21  & 0.07  & 29   \\ \hline
\cite{na61_hadrons}                   & 8.76  &  0.084 & 8    \\ \hline
\cite{gazdzicki:1996,antinucci:1973}  & 9.08  &  0.08  & 25  \\ \hline
\cite{gazdzicki:1996,antinucci:1973}  & 9.97  &  0.11  & 27   \\ \hline
\cite{gazdzicki:1996,antinucci:1973}  & 11.54 &   0.13  & 23   \\ \hline
\cite{na61_hadrons}                   & 12.32 &  0.095 & 7   \\ \hline
\cite{na61_hadrons}                   & 17.30 &  0.132 & 11   \\ \hline
\cite{na49_kaons}                     & 17.30 &  0.13  & 10   \\ \hline
\cite{antinucci:1973}                 & 22.02 & 0.24  & 10   \\ \hline
\cite{na49_kaons}                     & 23.00 & 0.171 & 15  \\ \hline
\cite{gazdzicki:1996}                 & 23.68 &  0.209 & 15   \\ \hline
\cite{gazdzicki:1996}                 & 30.59 &  0.244 & 15   \\ \hline
\cite{antinucci:1973,na49_kaons}      & 30.98 & 0.245 & 10  \\ \hline
 \end{tabular}\hspace{3mm}
         \begin{tabular}{|c|c|c|c|}
        \hline
        Reference & 
        $\sqrt{s_{NN}}$ (GeV) &
        $\left<K^+\right>$ & Error (\%) \\ \hline
\cite{gazdzicki:1996}                 & 2.98  &  0.0046  & 15  \\ \hline
\cite{gazdzicki:1996,antinucci:1973}  & 2.99  &  0.0035  & 16  \\ \hline
\cite{gazdzicki:1996}                 & 2.99  &  0.0044  & 18   \\ \hline
\cite{gazdzicki:1996}                 & 3.12  &  0.0057  & 18 \\ \hline
\cite{gazdzicki:1996}                 & 3.35  &  0.0069  & 15  \\ \hline
\cite{gazdzicki:1996,antinucci:1973}  & 3.50  &  0.008   & 21  \\ \hline
\cite{gazdzicki:1996}                 & 4.11  &  0.02    & 20  \\ \hline
\cite{gazdzicki:1996,antinucci:1973}  & 5.03  & 0.07    & 43   \\ \hline
\cite{gazdzicki:1996}                 & 5.35  & 0.054   & 10    \\ \hline
\cite{gazdzicki:1996,antinucci:1973}  & 6.15  & 0.107   & 2     \\ \hline
\cite{na61_hadrons}                   & 6.27  & 0.097   & 14   \\ \hline
\cite{gazdzicki:1996,antinucci:1973}  & 6.84  & 0.1188  &1 3  \\ \hline
\cite{na61_hadrons}                   & 7.75  & 0.157   & 12   \\ \hline
\cite{na61_hadrons}                   & 8.76  &  0.17    & 15   \\ \hline
\cite{gazdzicki:1996,antinucci:1973}  & 11.54 & 0.21    & 28  \\ \hline
\cite{na61_hadrons}                   & 12.32 & 0.201   & 7   \\ \hline
\cite{na61_hadrons}                   & 17.30 & 0.234   & 9   \\ \hline
\cite{na49_kaons}                     & 17.30 & 0.227   & 5    \\ \hline
\cite{antinucci:1973}                 & 22.02 & 0.35    & 10   \\ \hline
\cite{na49_kaons}                     & 23.00 &  0.273   & 15  \\ \hline
\cite{gazdzicki:1996}                 & 23.68 &  0.337   & 15   \\ \hline
\cite{gazdzicki:1996}                 & 30.59 &  0.367   & 15   \\ \hline
\cite{antinucci:1973,na49_kaons}      & 30.98 & 0.3562  & 13 \\ \hline

\end{tabular}}
     \caption{\label{table2}The compiled results on the mean
 multiplicity of charged kaons from inelastic proton-proton interactions at different collision energies.}
\end{center}
   \end{table}

In the framework of the NA61/SHINE experiment~\cite{na61}, the NA49 detector was upgraded
with a large solid angle TOF system of 100~ps resolution and faster trigger and read-out electronics.
The measurements of charged hadrons in $pp$ reactions were performed at five collision
energies: $\sqrt{s_{NN}}$\,=\,6.2, 7.6, 8.8, 12.3, and 17.3 GeV. The NA61 Collaboration published 
rapidity distributions and transverse momentum spectra for charged pions~\cite{na61_hminus,na61_hadrons},
kaons~\cite{na61_hadrons}, and (anti)protons~\cite{na61_hadrons}, covering about 1.5 rapidity
units in the forward region and the transverse momentum interval from 0 to 2 GeV/c.
The hadron yields were then extrapolated to the regions not covered by measurements and the
total multiplicities were obtained. Differential spectra, the rapidity distribution and the
total yield of $\Lambda$-hyperons were obtained only at the top SPS energy~\cite{na61_lambda}.

Experimental data on the mean multiplicity of $\pi^{\pm}$, K$^{\pm}$, K$^{0}_{S}$,
(anti)protons, and $\Lambda$ are tabulated in Tables~\ref{table1}-\ref{table4}.
The compilation includes the results in the energy range slightly wider than the one
for NICA: from about 3~GeV (above the threshold energy for strangeness production of~$\approx$\,2.6~GeV)
and up to 31 GeV.
In this paper $\Lambda$ denotes the sum of the average multiplicity of $\Lambda$-hyperons produced
directly and those originating from electromagnetic decays of $\Sigma^{0}$-hyperons.
The statistical and systematic errors of the measurements are added
in quadrature, and all the total errors are given in percent relative to the measured value in order
to simplify visual recognition of more(less) accurate data.

\begin{table}[htp]
 \begin{center}
 \scalebox{0.9}{%
 \small\addtolength{\tabcolsep}{-1pt}
        \begin{tabular}{|c|c|c|c|}
        \hline
        Reference & 
        $\sqrt{s_{NN}}$ (GeV) &
        $\left<K^0_S\right>$ & Error (\%) \\ \hline
 
\cite{gazdzicki:1996}                 & 2.98  &  0.00083 & 22 \\ \hline
\cite{gazdzicki:1996}                 & 3.35  & 0.0019  & 16 \\ \hline
\cite{gazdzicki:1996,antinucci:1973}  & 3.50  & 0.00364 & 3  \\ \hline
\cite{gazdzicki:1996}                 & 3.63  &  0.0034  & 9  \\ \hline
\cite{gazdzicki:1996}                 & 3.85  &  0.0064  & 8   \\ \hline
\cite{gazdzicki:1996}                 & 4.08  &  0.0072  & 8  \\ \hline
\cite{gazdzicki:1996,antinucci:1973}  & 4.93  & 0.0202  & 2   \\ \hline
\cite{gazdzicki:1996,antinucci:1973}  & 5.01  & 0.023   & 2 \\ \hline
\cite{gazdzicki:1996,antinucci:1973}  & 6.12  &0.0415  & 3   \\ \hline
\cite{gazdzicki:1996,antinucci:1973}  & 6.84  & 0.0495  & 2   \\ \hline
\cite{gazdzicki:1996}                 & 6.91  & 0.045   & 9   \\ \hline
\cite{gazdzicki:1996}                 & 11.45 &  0.109   & 6  \\ \hline
\cite{gazdzicki:1996}                 & 13.76 &  0.122   & 8   \\ \hline
\cite{gazdzicki:1996,antinucci:1973}  & 13.90 & 0.141   & 10  \\ \hline
\cite{gazdzicki:1996}                 & 16.66 & 0.158   & 4   \\ \hline
\cite{gazdzicki:1996}                 & 19.42 & 0.16    & 13  \\ \hline
\cite{gazdzicki:1996}                 & 19.66 &  0.181   & 8   \\ \hline
\cite{na49_kaons}                     & 23.00 &  0.222   & 10  \\ \hline
\cite{gazdzicki:1996}                 & 23.76 &  0.224   & 8    \\ \hline
\cite{gazdzicki:1996}                 & 26.02 &  0.26    & 4   \\ \hline
\cite{gazdzicki:1996}                 & 27.43 &   0.2     & 10  \\ \hline
\cite{gazdzicki:1996}                 & 27.60 &   0.232   & 5   \\ \hline
\cite{antinucci:1973,na49_kaons}      & 30.98 & 0.274   & 10  \\ \hline
      
\end{tabular}\hspace{3mm}
        \begin{tabular}{|c|c|c|c|}
        \hline
        Reference & 
        $\sqrt{s_{NN}}$ (GeV) &
        $\left<\Lambda\right>$ & Error (\%) \\ \hline
\cite{gazdzicki:1996}                            & 2.98  &  0.0033  & 18 \\ \hline
\cite{hades}                                     & 3.17  & 0.0073   & 4  \\ \hline
\cite{gazdzicki:1996}                            & 3.35  & 0.0073   & 4  \\ \hline
\cite{gazdzicki:1996,antinucci:1973}             & 3.50  & 0.0127   & 9  \\ \hline
\cite{gazdzicki:1996}                            & 3.63  &  0.0109   & 6  \\ \hline
\cite{gazdzicki:1996}                            & 3.85  & 0.0172   & 6  \\ \hline
\cite{gazdzicki:1996}                            & 4.08  &  0.0201   & 5  \\ \hline
\cite{gazdzicki:1996,antinucci:1973}             & 4.93  & 0.0388   & 2  \\ \hline
\cite{gazdzicki:1996,antinucci:1973}             & 5.01  &  0.035    & 11 \\ \hline
\cite{gazdzicki:1996,antinucci:1973}             & 6.12  & 0.061    & 3  \\ \hline
\cite{gazdzicki:1996,antinucci:1973}             & 6.84  & 0.0657   & 1  \\ \hline
\cite{gazdzicki:1996}                            & 6.91  & 0.037    & 19 \\ \hline
\cite{gazdzicki:1996}                            & 11.45 &  0.109   & 6 \\ \hline
\cite{gazdzicki:1996}                            & 13.76 & 0.112    & 12 \\ \hline
\cite{gazdzicki:1996,antinucci:1973}             & 13.90 & 0.099    & 12 \\ \hline
\cite{gazdzicki:1996}                            & 16.66 &  0.133    & 5  \\ \hline
\cite{na61_hadrons}                              & 17.30 &  0.12     & 9  \\ \hline
\cite{gazdzicki:1996}                            & 19.42 &  0.08     & 25 \\ \hline
\cite{gazdzicki:1996}                            & 19.66 &  0.103    & 11 \\ \hline
\cite{gazdzicki:1996}                            & 23.76 &  0.111    & 14 \\ \hline
\cite{gazdzicki:1996}                            & 23.76 &  0.11     & 9  \\ \hline
\cite{gazdzicki:1996}                            & 26.02 &  0.12     & 17 \\ \hline
\cite{gazdzicki:1996}                            & 27.43 &  0.12     & 8  \\ \hline
\cite{gazdzicki:1996}                            & 27.60 &  0.125    & 6  \\ \hline
\end{tabular}}
     \caption{\label{table3}The compiled results on the mean multiplicity of $\left<K^0_S\right>$
     and $\left<\Lambda\right>$ from inelastic proton-proton interactions at different collision
     energies.}
\end{center}
\end{table}

\begin{table}[htp]
 \begin{center}
 \scalebox{0.9}{%
 \small\addtolength{\tabcolsep}{-1pt}
        \begin{tabular}{|c|c|c|c|}
        \hline
        Reference & 
        $\sqrt{s_{NN}}$ (GeV) &
        $\left<p\right>$ & Error (\%) \\ \hline

\cite{gazdzicki:1996,antinucci:1973}             & 3.50  & 1.56  & 10  \\ \hline
\cite{gazdzicki:1996,antinucci:1973}             & 4.93  &  1.68  & 10  \\ \hline
\cite{gazdzicki:1996,antinucci:1973}             & 5.01  &  1.55  & 10  \\ \hline
\cite{gazdzicki:1996,antinucci:1973}             & 6.12  & 1.41  & 10  \\ \hline
\cite{gazdzicki:1996,antinucci:1973}             & 6.15  & 1.69  & 10  \\ \hline
\cite{na61_hadrons}                              & 6.27  &  1.154 & 4 \\ \hline
\cite{gazdzicki:1996,antinucci:1973}             & 6.84  &  1.615 & 10 \\ \hline
\cite{na61_hadrons}                              & 7.75  & 1.093 & 6  \\ \hline
\cite{na61_hadrons}                              & 8.76  &  1.095 & 8  \\ \hline
\cite{na61_hadrons}                              & 12.32 &  0.977 & 14  \\ \hline
\cite{na61_hadrons}                              & 17.30 & 1.069 & 12  \\ \hline
\cite{na49_kaons}                                & 17.30 &  1.162 & 15  \\ \hline
\cite{antinucci:1973}                            & 22.02 &  1.28  & 10  \\ \hline
\cite{antinucci:1973,na49_kaons}                 & 30.98 &  1.34  & 10   \\ \hline      
\end{tabular}\hspace{3mm}
        \begin{tabular}{|c|c|c|c|}
        \hline
        Reference & 
        $\sqrt{s_{NN}}$ (GeV) &
        $\left<\bar{p}\right>$ & Error (\%) \\ \hline

\cite{gazdzicki:1996,antinucci:1973}             & 6.15  & 0.0023 & 10  \\ \hline
\cite{na61_hadrons}                              & 6.27  & 0.0047 & 15 \\ \hline
\cite{gazdzicki:1996,antinucci:1973}             & 6.84  & 0.004  & 10  \\ \hline
\cite{na61_hadrons}                              & 7.75  & 0.0047 & 16  \\ \hline
\cite{gazdzicki:1996,antinucci:1973}             & 8.21  & 0.005  & 10 \\ \hline
\cite{na61_hadrons}                              & 8.76 & 0.0059 & 12 \\ \hline
\cite{gazdzicki:1996,antinucci:1973}             & 9.08  & 0.008  & 10  \\ \hline
\cite{gazdzicki:1996,antinucci:1973}             & 9.97  & 0.011  & 10  \\ \hline
\cite{gazdzicki:1996,antinucci:1973}             & 11.54 & 0.015  & 10  \\ \hline
\cite{na61_hadrons}                              & 12.32 & 0.0183 & 10  \\ \hline
\cite{na61_hadrons}                              & 17.30 & 0.0402 & 9   \\ \hline
\cite{na49_kaons}                                & 17.30 & 0.039  & 15  \\ \hline
\cite{antinucci:1973}                            & 22.02 & 0.061  & 10  \\ \hline
\cite{antinucci:1973,na49_kaons}                 & 30.98 & 0.11   & 10  \\ \hline
\end{tabular}}
     \caption{\label{table4}The compiled results on the mean multiplicity of $\left<p\right>$ and $\left<\bar{p}\right>$    
from inelastic proton-proton interactions at different collision energies.}
\end{center}
\end{table}
        
In addition to that, in Table~\ref{table5} are tabulated the results for $\pi^{-}$ obtained
from the measurements of negatively charged hadrons ($h^-$) from~\cite{gazdzicki:1995}.
The contribution from K$^{-}$ and antiprotons in the $h^-$ yields was subtracted by us,
the correction procedure is described in Section~\ref{sec_pions}.
 
Special emphasis should be put here on the problem of the contribution from weak decays of strange
particles in the measured multiplicities of pions and (anti)protons. The size of this contribution
depends on the collision energy and used experimental technique. In bubble chamber
experiments of a small detector size most weak decays escape the fiducial volume and the
remaining secondary decay vertices can be easily reconstructed, thus the daughters are eliminated from
the analysis. In contrast, for fixed target experiments with long detectors a sizeable fraction
of the weak decay products populate the data sample due to limited vertex finding performance in
such setups. In the data collection, presented here, there is no unique approach to deal
with the feeddown correction: in~\cite{na49_pions,na49_ppbar} the contribution from weak decays
was subtracted, in~\cite{na61_hadrons} it was partially eliminated by track selection criteria, while
in many fixed target experiments from the last century decay products partially counted
into the total sample without any attempt to make a correction for their contribution.
As a reference estimate of the size of such a correction we can refer to~\cite{na49_pions},
where a complete feed-down subtraction procedure for charged pions from $pp$ reactions
at $\sqrt{s_{NN}}=17.3$~GeV was performed by considering all relevant sources
(i.e. $K^0_S$, $\Lambda$, $\Sigma^0$, $\Sigma^{\pm}$). The authors found that the feed-down pions
are mostly concentrated at low $p_t$ near midrapidity having percentage of 6\% for $\pi^+$
and 10\% for $\pi^-$. Of course, the overall (integrated over the full phase space) feed-down
contribution is smaller, nevertheless, an additional systematic uncertainty to the results
on pion multiplicities reported by early experiments (of the order of 3-5\%) should be kept in mind.

\begin{table}[htp]
\begin{center}
\scalebox{0.9}{%
 \small\addtolength{\tabcolsep}{-4pt}
  \begin{tabular}{|c|c|c|c|}
 \hline
        Reference & $\sqrt{S_{NN}}$ (GeV) &
         $\left<\pi^-\right>$ ($\left<h^-\right>$) & Error (\%)  \\ \hline
\cite{gazdzicki:1995}& 2.98   & 0.170 & 14  \\ \hline
\cite{gazdzicki:1995}& 3.5 & 0.352 & 3 \\ \hline
\cite{gazdzicki:1995}& 3.78 & 0.429 & 5\\ \hline 
\cite{gazdzicki:1995}& 4.93 & 0.707 & 2\\ \hline
\cite{gazdzicki:1995}& 6.12 & 0.985 & 2 \\ \hline
\cite{gazdzicki:1995}& 6.84 & 1.08 & 2 \\ \hline
\cite{gazdzicki:1995}& 8.29 & 1.33 & 2\\ \hline 
\cite{gazdzicki:1995}& 9.78 & 1.60 & 4 \\ \hline
\cite{gazdzicki:1995}& 10.69 & 1.705 & 4 \\ \hline
\cite{gazdzicki:1995}& 11.46 & 1.82 & 3 \\ \hline
\cite{gazdzicki:1995}& 13.76 & 2.07 & 2\\ \hline 
\cite{gazdzicki:1995}& 13.9 & 2.02 & 3 \\ \hline
\cite{gazdzicki:1995}& 16.66 & 2.34 & 3 \\ \hline
\cite{gazdzicki:1995}& 18.17 & 2.57 & 5\\ \hline 
\cite{gazdzicki:1995}& 19.42 & 2.65 & 3\\ \hline
\cite{gazdzicki:1995}& 19.66 & 2.63 & 3 \\ \hline
\cite{gazdzicki:1995}& 21.7 & 2.72 & 4 \\ \hline
\cite{gazdzicki:1995}& 23.88 & 3.09 & 3\\ \hline
\cite{gazdzicki:1995}& 26.02 & 3.26 & 3 \\ \hline

\end{tabular}}

\caption{\label{table5}The results on mean multiplicities $\left<\pi^{-}\right>$
calculated from the data for negatively charged hadrons $\left<h^-\right>$~\cite{gazdzicki:1995}
in inelastic proton-proton interactions at different collision energies. The calculation
procedure is described in the text.}
\end{center}
\end{table}
 
\section{Study of the excitation function of the mean multiplicity of hadrons}
\label{sec2}

\hspace{4mm}In the following subsections we discuss the excitation functions of hadron yields
 specie by specie.

\subsection{Kaons}
\label{sec_kaons}

\begin{figure}[hbt]
  \centerline{
    \includegraphics[width=85mm,height=60mm,angle=0]{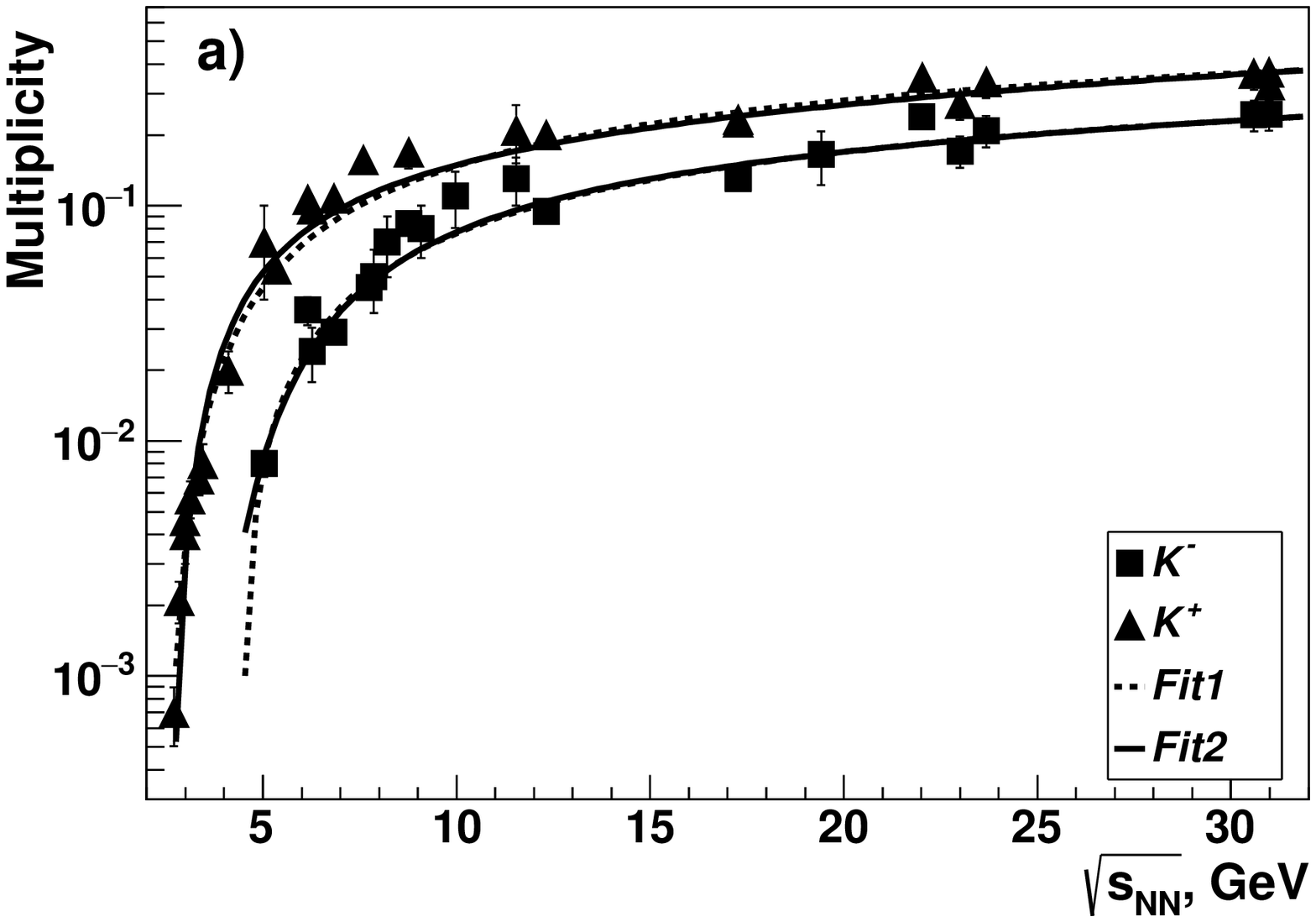}
    \includegraphics[width=85mm,height=60mm,angle=0]{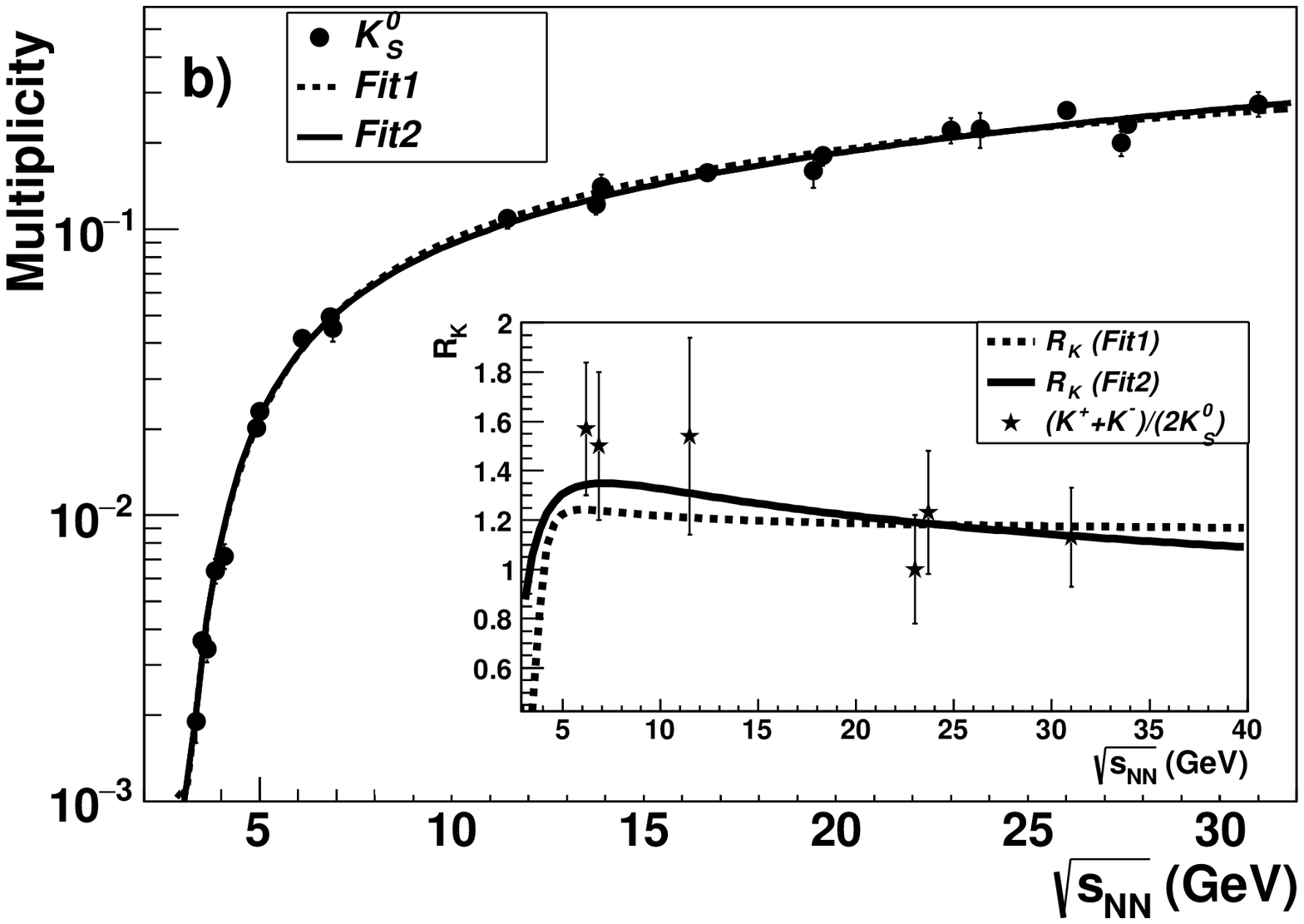}
}

\caption{a) The energy dependence of the multiplicity of charged kaons
 in inelastic $pp$ collisions (data from Table~\ref{table2}).
 Fits to Eq.\ref{eq1} and Eq.\ref{eq2} are shown by dashed and solid lines, respectively
 (see text for details).  
b) The multiplicity of K$^0_S$ as a function of $\sqrt{s_{NN}}$
 in inelastic $pp$ collisions (data from Table~\ref{table3}).
 Fits to Eq.\ref{eq1} and Eq.\ref{eq2} are shown by dashed and solid lines, respectively.
  In the inset the $R_K$-ratio is plotted
 as a function of center-of-mass energy for the two discussed parameterizations 
 (dashed and solid lines) and for kaon measurements (symbols).
}
 \label{mult_kaons}
\end{figure}

\hspace{4mm} We begin with kaons because these results are used in Sec.~\ref{sec_pions}
to obtain the yields of $\pi^-$ from those of negatively charged hadrons. 
The energy dependence of the mean multiplicity of charged
kaons $\left<K^{\pm}\right>$ from minimum-bias $pp$ collisions is shown in Fig.\,\ref{mult_kaons}a.
A strong dependence of the kaon production rates on $\sqrt{s_{NN}}$ is observed  at low NICA energies
due to threshold-dominated  effects.
In order to describe the excitation function of the hadron production rates we used
two parameterizations. The first one ({\it Fit1}) represents a fit to the form

\begin{equation}
\label{eq1}
\left< n\right> = a + \frac{b}{\sqrt{s}} + c\ln{s}
\end{equation}

with 3 parameters (a,b,c), which was suggested in~\cite{mult_redge} based on a general
analysis of hadron multiplicities including Redge trajectory with intercept one-half. 

In addition, we used a parameterization based on the Lund-String-Model (LSM) from~\cite{cass_brat}
({\it Fit2}) given by


\begin{equation}
\label{eq2}
\left< n\right> = a\left(x-1\right)^b\left(x\right)^{-c}
\end{equation}

where $x=s/s_0$, $s$ is the square of the center-of-mass energy, $s_0$ is the square of the 
production threshold, and ($a,b,c$) are the fit parameters.

The fit results are plotted in Fig.~\ref{mult_kaons}a with a dashed and solid line
for {\it{Fit1}} and {\it{Fit2}}, respectively. 
We found that both parameterizations describe the energy dependence of hadron
multiplicities through the NICA energy range equally good, while {\it{Fit2}} is slightly better
describes the overall trend (see Fig.~\ref{mult_kaons}b and Fig.~\ref{mult_pions}).

As it was underlined in the Introduction, it is interesting to look at the available data on $K^0_S$
production, because the relation between charged and neutral kaons is also strongly related to the
competitive strangeness production mechanisms.  The energy dependence for the mean multiplicity
of $K^0_S$ from $pp$ collisions is shown in Fig.~\ref{mult_kaons}b.
Fits to Eq.\ref{eq1} and Eq.\ref{eq2} are shown by dashed
and solid lines, respectively. Because of isospin conservation in strong
interactions, the production of charged and neutral kaons species is equally probable, thus,
from isospin invariance one expects the ratio
$$
R_K=\frac{0.5(\left<n_{K^+}\right> + \left<n_{K^-}\right>)}{\left<n_{K^0_S}\right>}
$$
to be equal to unity (only half of the produced neutral kaons is considered as $K^0_L$
is difficult to register due to its long lifetime).

The inset in Fig.~\ref{mult_kaons}b indicates the energy dependence for
$R_K$ taken as the ratio of the parametrized excitation functions of kaon multiplicities
for {\it Fit1} (dashed line) and {\it Fit2} (solid line). Moreover, among all the collected data
points we found those where all kaon sorts (i.e. $K^+,K^-,K^0_S$) are measured
at the same energy or at very closed energies. Thus, the $R_K$ ratio was calculated from this subset
of measurements and shown in the inset of Fig.~\ref{mult_kaons}b by stars.
As can be seen, $R_K$ approaches unity with increasing energy, still, deviating from the isospin
invariance motivated kaon production ratio in the NICA energy range.
We also found that the results for {\it Fit2} slightly better describe the trend for $R_K$.
Since the experimental errors in the discussed kaon measurements are large, we can not argue that
the observed deviation in the $R_K$-ratio from the expected value might be an indicator
of some anomaly in the kaon production in $pp$ interactions at NICA energies 
(as non-trivial interference of multiple isospin production channels or a signal
for isospin fluctuations in kaon sector). New high precision measurements for
all kaon species in the NICA energy range can potentially unveil this mystery.  
 


One advantage concerning the kaon data is the absence of feed-down from weak decays, 
except the contribution from decays of $\Omega$-hyperons, which is negligibly small at NICA energies.
From other side, the contribution from decays of resonances (as $\phi(1020)$) needs to be (at least)
estimated. Avoiding the region of collision energies close to the production threshold, the ratio
$\left<\phi\right>/\left<\pi\right>\approx0.004$
can be obtained from a number of measurements in the NICA energy range~\cite{na49_phi}. Taking
into account the averaged charged pion yield and the branching to charged and neutral kaons
we estimate at the top NICA energy the $\phi-$meson contribution to the production rates of kaons
as: 2.1\% for $K^{+}$, 3.6\% for $K^{-}$, and $<$1\% for $K^0_S$. These numbers are well below
the errors quoted in the data compilation, so we neglect this contribution.  

  
\subsection{(Anti)protons and $\Lambda$-hyperons}
\label{sec_prot}
\hspace{4mm}Experimental data for the mean multiplicity of protons and antiprotons from $pp$
interactions in the NICA energy range are scarce. In the case of antiprotons the reason is the
low production cross section, thus very large data sets have to be collected to provide
sufficient statistics. However, the experimental technique in old (mostly bubble chamber) experiments
did not allow to collect large volume data sets, only very recently such measurements became
available from SPS~\cite{na49_ppbar,na61_hadrons}.
For the case of protons the reason of small number of inclusive measurements is different.
Since protons in $pp$ reactions at NICA are mainly produced from fragmentation of strings
(with small addition from decays of $\Delta$ resonances), their distributions in rapidity
have a non-trivial shape (see Fig.~\ref{rap_pika}b). Since measurements  over a limited acceptance
does allow to do easy extrapolation to unmeasured regions, only experiments with a wide phase space
coverage can provide valuable data for protons (actually, a close to $2\pi$-coverage in the
forward/backward direction is sufficient). 
     
\begin{figure}[hbt]
  \centerline{
    \includegraphics[width=85mm,height=60mm,angle=0]{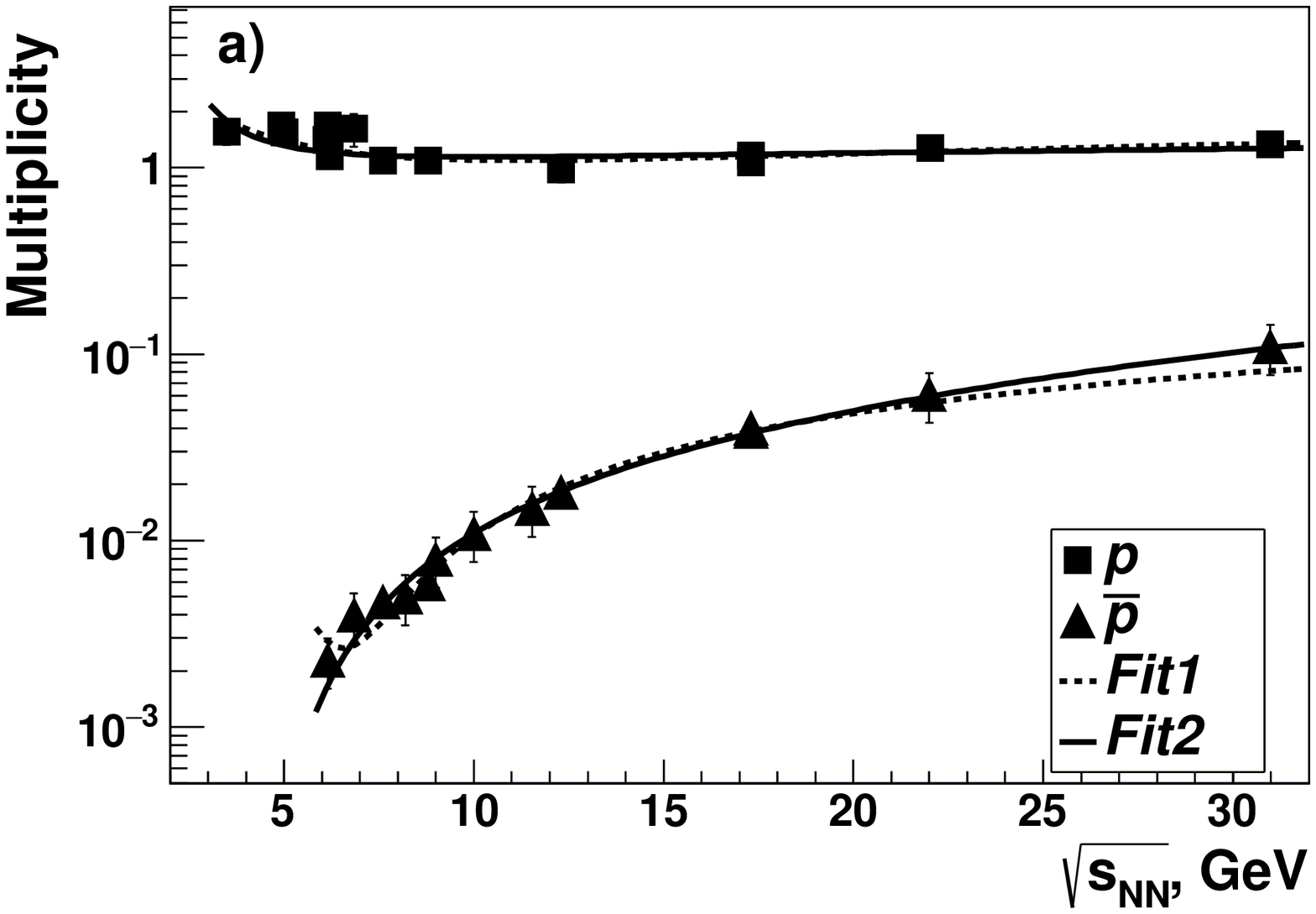}
    \includegraphics[width=85mm,height=60mm,angle=0]{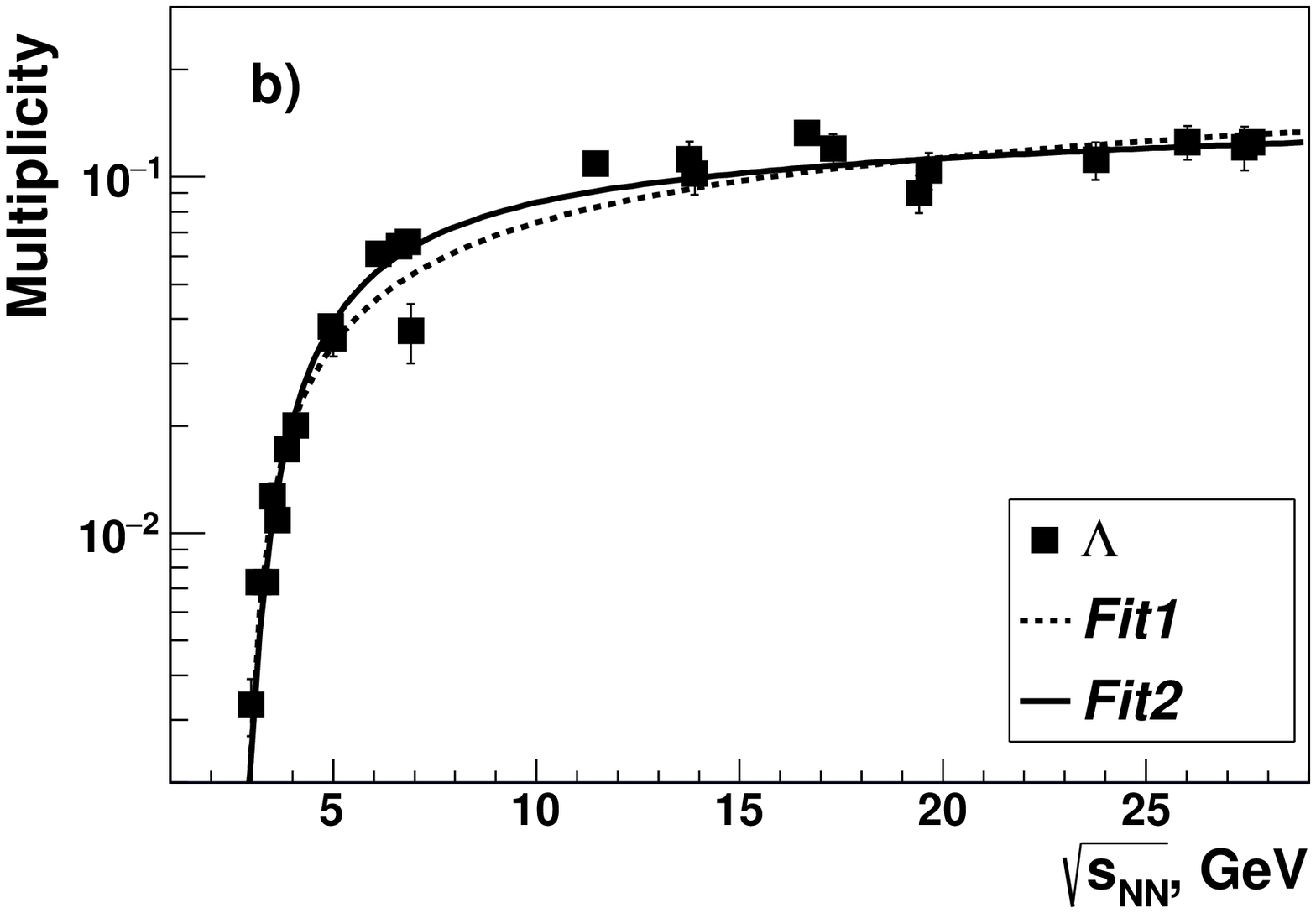}
}

\caption{a) The energy dependence of the multiplicity of charged protons and antiprotons
 in inelastic $pp$ collisions (data from Table~\ref{table3}). Fits to Eq.\ref{eq1}
  and Eq.\ref{eq2} are shown by dashed and solid lines, respectively.  
b) The energy dependence of the multiplicity of $\Lambda$ in inelastic $pp$ collisions (data
 are from Table~\ref{table4}).
}
 \label{mult_protons}
\end{figure}

The energy dependence for the mean multiplicity of protons and antiprotons
is shown in Fig.~\ref{mult_protons}a. There is a small variation of the proton's multiplicity
in the NICA energy range, which is expected for a leading particle. The local minimum of
the yield of protons is reached at about 10 GeV and further increase can be roughly
accounted by the gain in the number of the produced baryon-antibaryon pairs. 
In contrast, the production of antiprotons grows rapidly at low NICA energies, exhibiting
a kind of threshold behavior up to large $\sqrt{s}$ values.

Measurements of $\Lambda$-hyperon production in the energy range of NICA are mainly performed
by bubble chamber experiments and only a single data point was obtained recently by the SPS NA61/SHINE
experiment at $\sqrt{s_{NN}}$\,=\,17.3~GeV~\cite{na61_lambda}.
The results for $\Lambda$-hyperons are presented in Fig.~\ref{mult_protons}b.
After a rapid increase close to the threshold, the $\Lambda$ production rate
is close to saturation at NICA energies. 
  

\subsection{Charged pions}
\label{sec_pions}

\begin{figure}[hbt]
  \centerline{
    \includegraphics[width=85mm,height=60mm,angle=0]{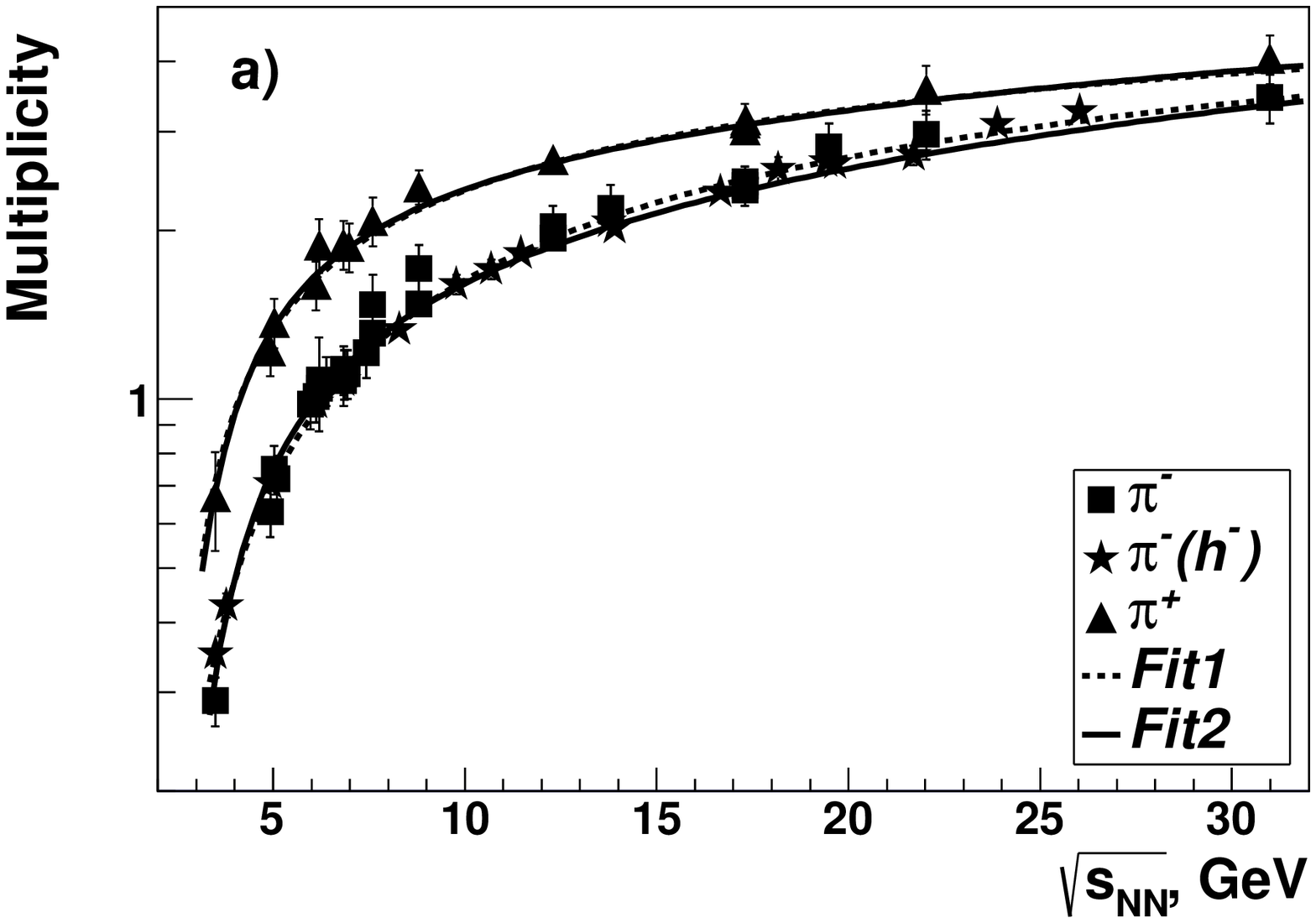}
   \includegraphics[width=85mm,height=60mm,angle=0]{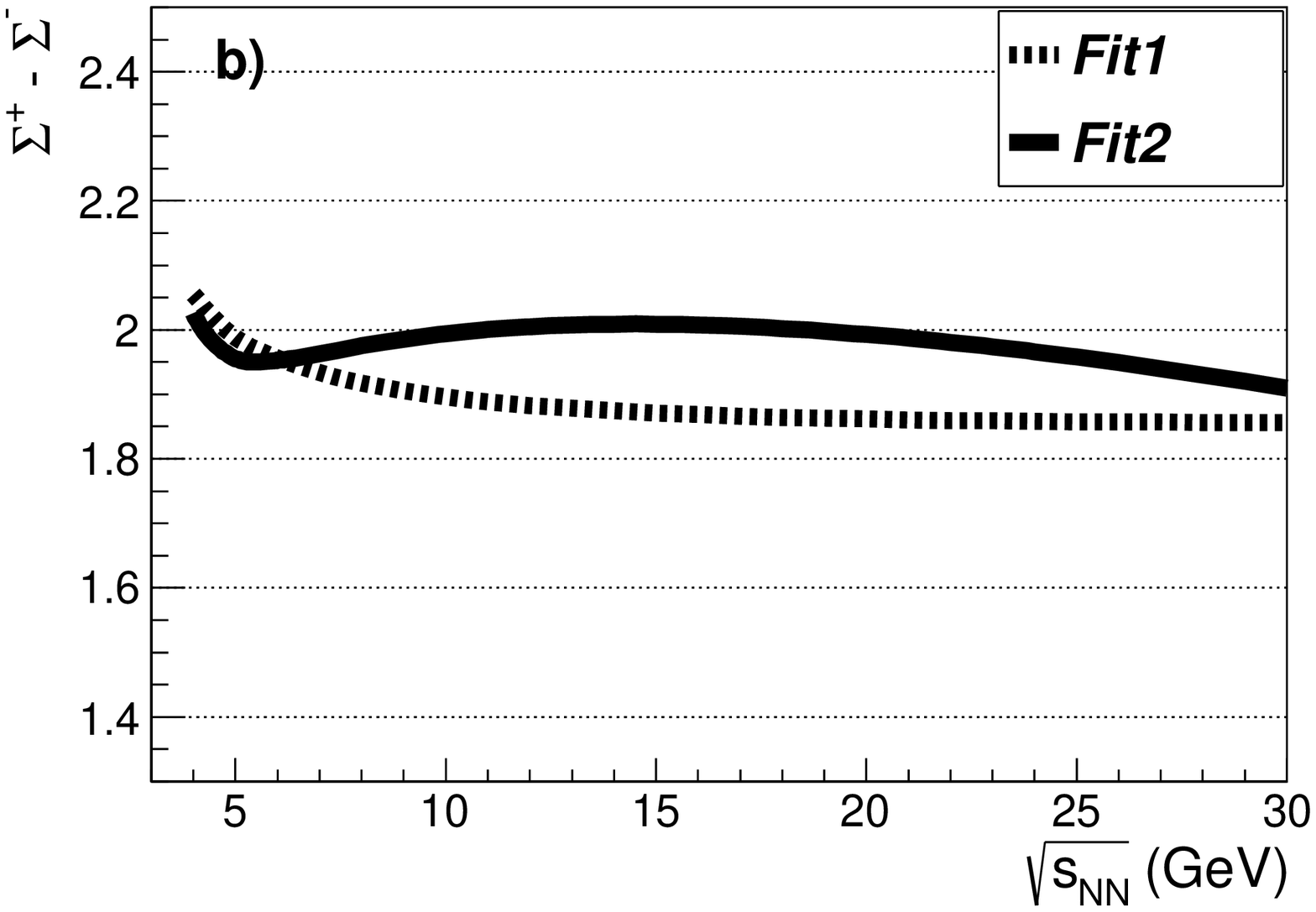}
}

\caption{a) The energy dependence of the multiplicity of charged pions
 in inelastic $pp$ collisions (data from Tables~\ref{table1},\ref{table5}).
 Fits to Eq.\ref{eq1}
  and Eq.\ref{eq2} are shown by dashed and solid lines, respectively.
b)The sum of the multiplicities of positively charged hadrons $\Sigma^+$ minus $\Sigma^-$
for all negatively charged ones (taken from the parameterizations for the hadron excitation
functions) as a function of $\sqrt{s_{NN}}$ (see text for details).}
 \label{mult_pions}
\end{figure}

\hspace{4mm}Now we come to the most abundant species - charged pions. In Figure~\ref{mult_pions}a
the mean multiplicity of charged pions in $pp$ interactions is plotted as a function of the collision
energy. The experimental data for $\pi^-$ and $\pi^+$ are drawn by squares and triangles, respectively.
In order to get the yield of $\pi^-$ from the measurements of negatively charged
hadrons from~\cite{gazdzicki:1995} the following procedure was used.
For each data point (for each $\sqrt{s_{NN}}$ value) the mean multiplicities of $K^-$ and antiprotons
from the known parameterizations for their energy dependences (see Sections~\ref{sec_kaons}
and~\ref{sec_prot}) were subtracted.
For this we used the results for the parameterization with {\it Fit2}. 
The uncertainty of this correction (calculated by propagating the fit errors) was added in quadrature
to the measurement error. The resulting negatively charged pion multiplicities are drawn
in Fig.~\ref{mult_pions}a by stars. The drawn in figure lines are the parameterizations
according to Eq.\ref{eq1} and Eq.\ref{eq2}.

The overall quality of the obtained parameterizations for hadrons can be tested if the excitation function
of the charge balance will be obtained. Conservation of charge in $pp$ reactions requires that
the difference between positive and negative particle multiplicities should give two units.
In Fig.~\ref{mult_pions}b the difference between the sum of the multiplicities for all positively
charged hadrons $\sum\nolimits^{+}= \left<n_{\pi^+}\right> +\left<n_{K^+}\right> +\left<n_{p}\right>$
and negatively charged hadrons
$\sum\nolimits^{-}= \left<n_{\pi^-}\right> +\left<n_{K^-}\right> +\left<n_{\bar{p}}\right>$ is plotted.
All the numbers are obtained from the two used parameterizations of data points drawn
in Figs.~\ref{mult_kaons}-\ref{mult_pions}.
The difference $\sum\nolimits^{+}-\sum\nolimits^{-}$ is also
accounts for the charge balance of charged hyperons
$\left<n_{\Sigma^+}\right> - \left<n_{\Sigma^-}\right>\approx 0.23\cdot\left<n_{\Lambda+\Sigma^0}\right>$,
the coefficient 0.23 was obtained from models.
We found that the overall charge balance is off from the nominal value (=2) not larger than
by 0.13 units, so all the multiplicities computed from the obtained parameterizations
are consistent with charge conservation within 5.9\% and 1.5\% for {\it Fit1} and {\it Fit2},
respectively.

\begin{figure}[hbt]
  \centerline{
   \includegraphics[width=0.65\textwidth]{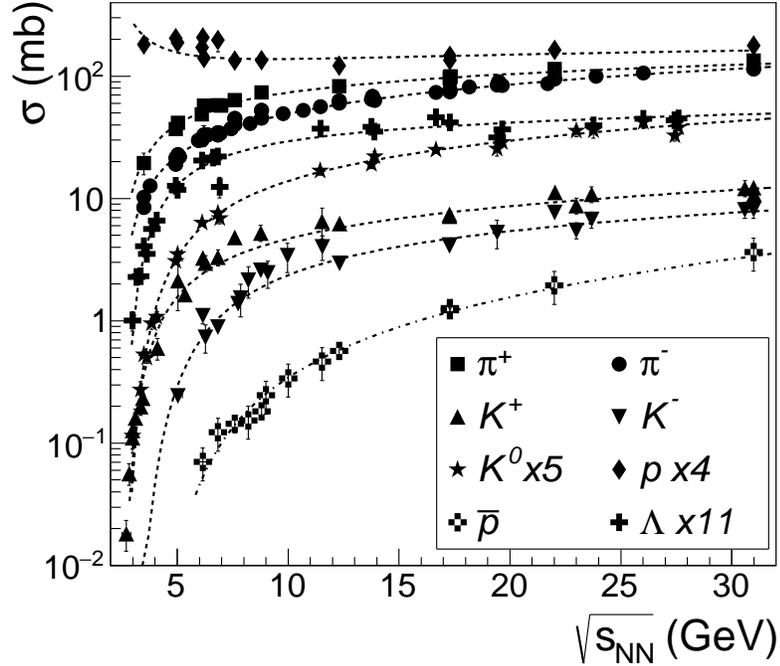}
}

\caption{\label{hadr_cross} Inclusive hadron production cross-sections from $pp$ interactions
as a function of the center-of-mass energy. Dashed lines are parameterizations to Eq.~\ref{eq2}.
}
\end{figure}

Finally, the inclusive production cross-sections for $\pi^{\pm}$, K$^{\pm}$, K$^0_S$, $\Lambda$, and
(anti)protons are drawn in Fig.~\ref{hadr_cross}.
They are calculated as $\sigma=\left<n\right>\sigma_{in}$,
where $\left<n\right>$ is the hadron multiplicity and  $\sigma_{in}$ is the inelastic cross-section.
The latter was calculated according the parameterization from~\cite{cross_sect}.
We used the functional form according to Eq.\ref{eq2} to parametrize the energy dependence for the
cross-sections and the fit parameters of the plotted lines  for all hadrons are tabulated
in Table~\ref{table8}.
 As one can see,
 $\chi^2/NDF$ lies in the range from 0.5 to 2.4 for all hadrons, except $\Lambda$.
 In the latter case, the goodness of the fit
can be evaluated visually (see Fig.~\ref{mult_protons}b), while a greater value of $\chi^2$ can
be explained by larger scattering of data points around the best fit line and by a couple of outliers
(i.e. measurements deviating more than 3$\sigma$ from the prediction).

\begin{table}[htp]
\begin{center}
\scalebox{0.9}{%
 \small\addtolength{\tabcolsep}{-4pt}
  \begin{tabular}{|c|c|c|c|c|c|}
        \hline

        Hadron & a & b & c & $s_0 (GeV^2)$ & $\chi^{2}$/NDF \\ \hline
$\pi^-$ & $18.79\pm 0.554$ & $1.998\pm 0.089$ & $-.653\pm 0.095$ & 4.64 & 0.5\\ \hline
$\pi^+$ & $43.046\pm 13.69$ & $2.366\pm 1.386$ & $2.168\pm 1.454$ & 4.07 & 0.5\\ \hline
$K^-$ & $1.509\pm 0.363$ & $5.138\pm 0.801$ & $4.783\pm 0.853$ & 8.2 & 2.0\\ \hline
$K^+$ & $2.176\pm 0.26$ & $2.63\pm 0.155$ & $2.285\pm 0.181$ & 6.49 & 2.4\\ \hline
$K^0_S$ & $1.151\pm 0.087$ & $3.697\pm 0.122$ & $3.284\pm 0.139$ & 6.49 & 1.9\\ \hline
$p$ & $19.49\pm 1.824$ & $-8.717\pm 0.054$ & $-8.823\pm 0.054$ & 0 & 1.3\\ \hline
$\bar{p}$ & $0.122\pm 0.004$ & $3.511\pm 0.291$ & $2.69\pm 0.271$ & 14.08 & 0.8\\ \hline
$\Lambda$ & $2.066\pm 0.161$ & $2.625\pm 0.102$ & $2.468\pm 0.121$ & 6.49 & 4.1\\ \hline
\end{tabular}}
\caption{\label{table8}Parameterization parameters (according to Eq.\ref{eq2}) for the 
hadron production cross-sections}
\end{center}
\end{table}

\section{Rapidity and transverse momentum distributions for hadrons from $pp$ reactions}
\label{sec4}

\begin{figure}[hbt]
  \centerline{
    \includegraphics[width=85mm,height=58mm,angle=0]{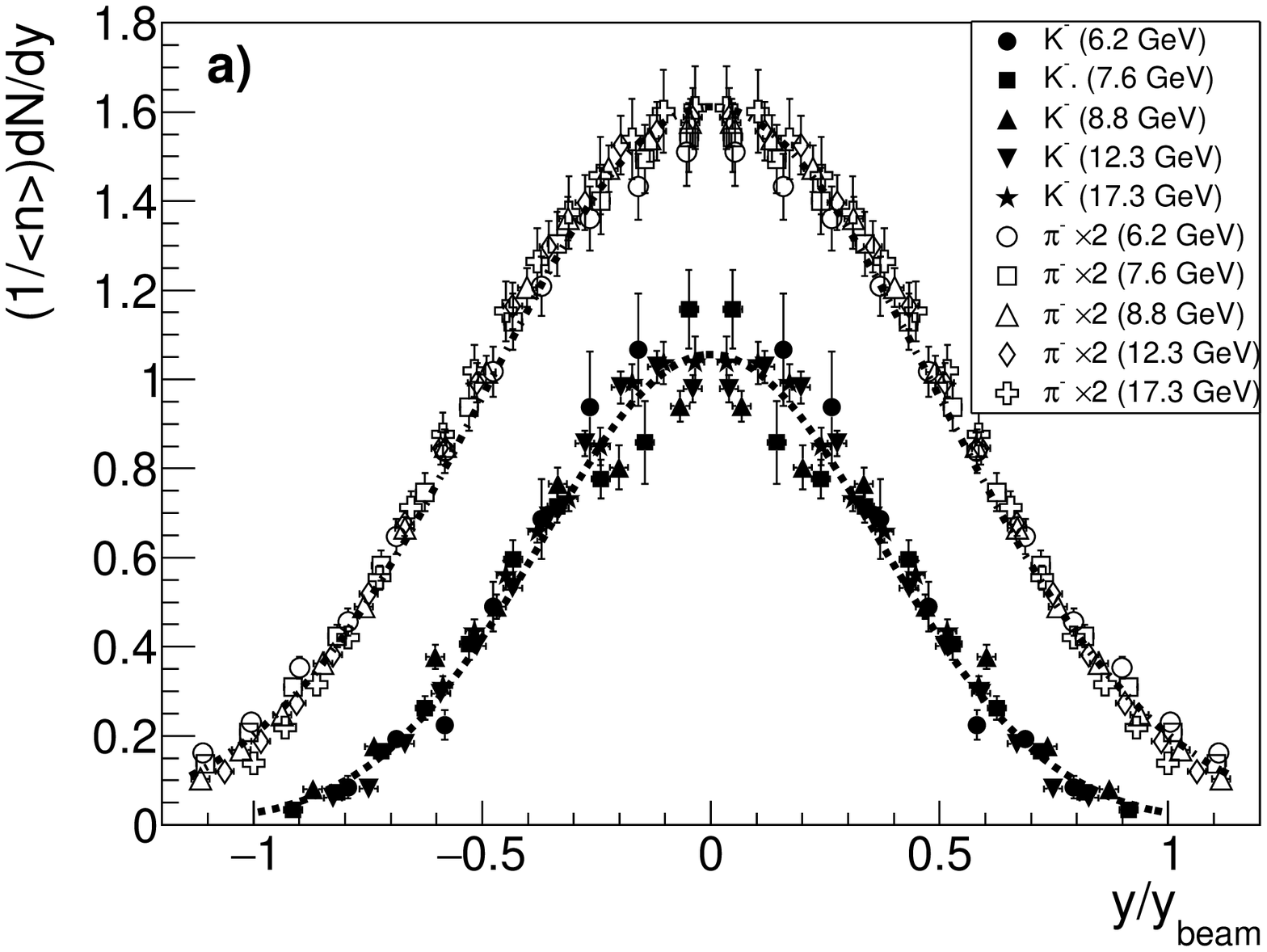}
    \includegraphics[width=85mm,height=60mm,angle=0]{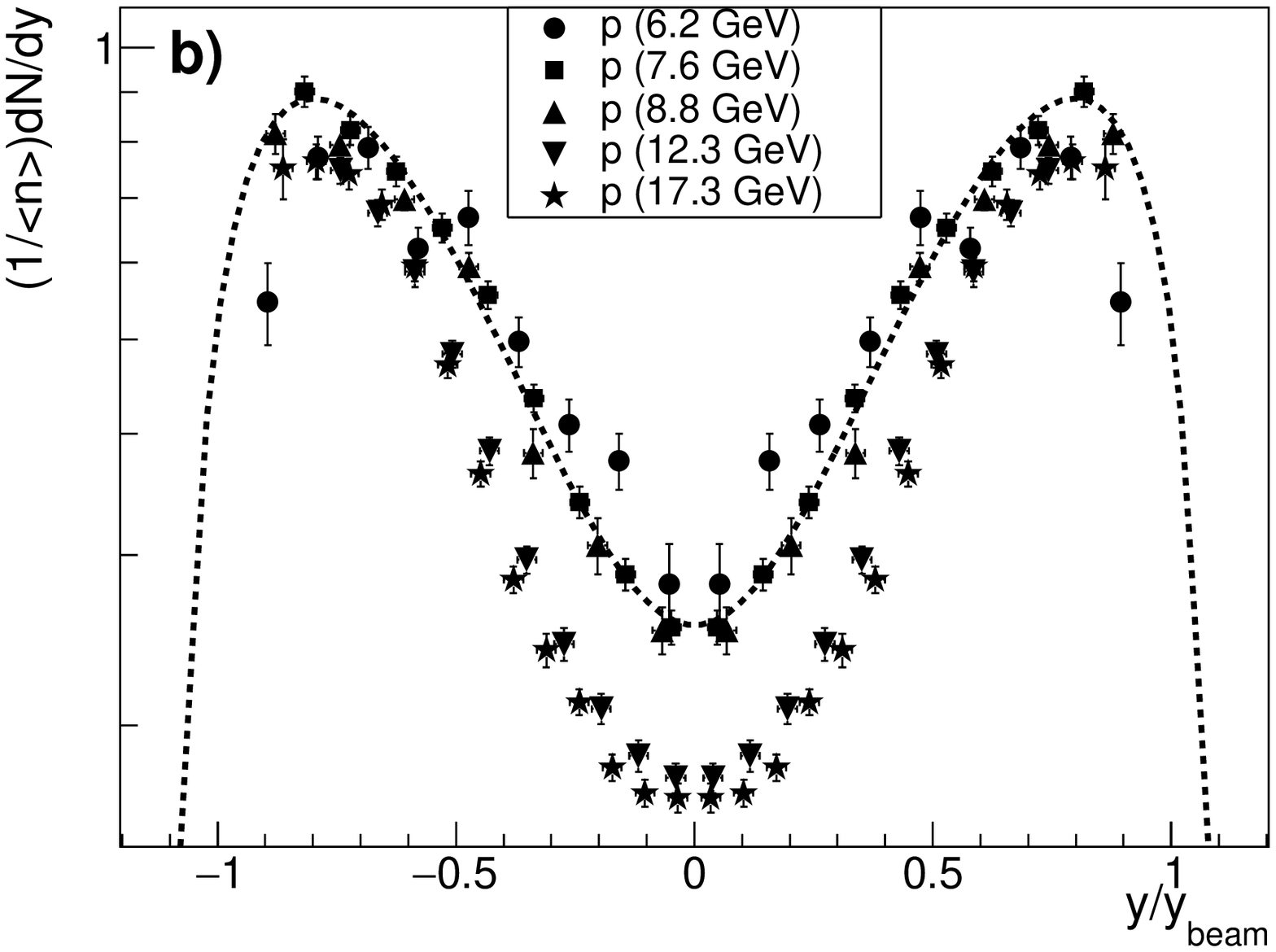}
}

\caption{a) The scaled yield of K$^-$ and $\pi^-$ as a function of normalized rapidity $y/y_{beam}$
 in inelastic $pp$ collisions at 6.2-17.3 GeV. The experimental data are taken from~\cite{na61_hadrons}
 and \cite{na61_hminus} and center-of-mass collision energy is given in the parenthesis.
 The numbers for $\pi^-$ are multiplied by 2 for clarity, the dashed lines indicate fits to Gaussian
 function (see text for details).  
b) The scaled yield of protons as a function of normalized rapidity $y/y_{beam}$
 in inelastic $pp$ collisions at 6.2-17.3 GeV. The experimental data are taken from~\cite{na61_hadrons}
 and center-of-mass collision energy is given in the parenthesis. The dashed line indicate a fit
  to the data at 7.6 GeV with a symmetric function of 6$^{th}$ order (see text for details)}
 \label{rap_pika}
\end{figure}

\hspace{3mm}In this section we discuss rapidity and transverse momentum distributions of hadrons
from $pp$ collisions. In Fig.~\ref{rap_pika}a rapidity distributions for negatively
charged kaons and pions are shown. These data from the NA61 experiment~\cite{na61_hadrons,na61_hminus}
were taken at five collision energies ($\sqrt{s_{NN}}$ from 6 to 17~GeV).
The data points were normalized by us to the mean multiplicity
$\left<n\right>$ and plotted as a function of the scaled rapidity $y$/$y_{beam}$,
where $y_{beam}$ is the rapidity value for the projectile proton.
Since the discussed measurements are only performed in the forward hemisphere ($y/y_{beam}>0$),
we have added the points in the backward hemisphere due to reflection symmetry with respect
to the midrapidty in $pp$ reactions. As can be seen, the rapidity dependence of the yields of pions
and kaons follows a bell-like shape at all energies. Moreover, the normalized distributions have
little shape variation within the energy interval $6<\sqrt{s_{NN}}<17$~GeV. Thus, Gaussian fits
were applied to the collection of rapidity spectra at all energies keeping the position of the
maximum fixed at midrapidity. The resulting fit functions are plotted by dashed lines.
The same procedure was applied to positively charged pions and kaons (not shown here) and the
parameter $\sigma$ of the Gaussian function give us the value of 0.37, 0.49, 0.51, and 0.55
for K$^-$, $\pi^-$, K$^+$, and $\pi^+$, respectively.

In Fig.~\ref{rap_pika}b scaled by the mean multiplicity rapidity distributions for protons
from minimum bias $pp$ reactions at five collision energies are shown.
The data points are taken from Ref.~\cite{na61_hadrons} and the measurements at forward rapidities
are reflected to the backward hemisphere assuming symmetry of the distribution around midrapidity. 
As one can see, rapidity spectra of protons in contrast to mesons and antiprotons have a broad
minimum at midrapidity rising rapidly towards beam (target) rapidity. Moreover,
the dip of the minimum increases with collision energy indicating increase of the rapidity loss
by initial protons in the course of the reaction, leading to the increase of the 
rapidity density of particles (mostly mesons) produced near midrapidity. We found that the shape
of rapidity spectra for protons can be parametrized by a symmetric polynomial of 6$^{th}$ order
$d$N/$d$y$\approx a(y/y_{beam})^6+b(y/y_{beam})^2+c$, where the coefficients of the parabolic part
of the fit function($b$ and $c$) describe the behavior near midrapidity and normalization,
while a negative coefficient $a$ defines sharp drop of the rapidity spectra at the beam rapidity.
An example of a fit with a symmetric polynomial of 6$^{th}$ order to the data at 7.6 GeV is drawn
in Fig.~\ref{rap_pika}b as a dashed line. The same parameterization can describe the rapidity
spectrum of $\Lambda$-hyperons (not shown).

\begin{figure}[hbt]
  \centerline{
    \includegraphics[width=0.65\textwidth]{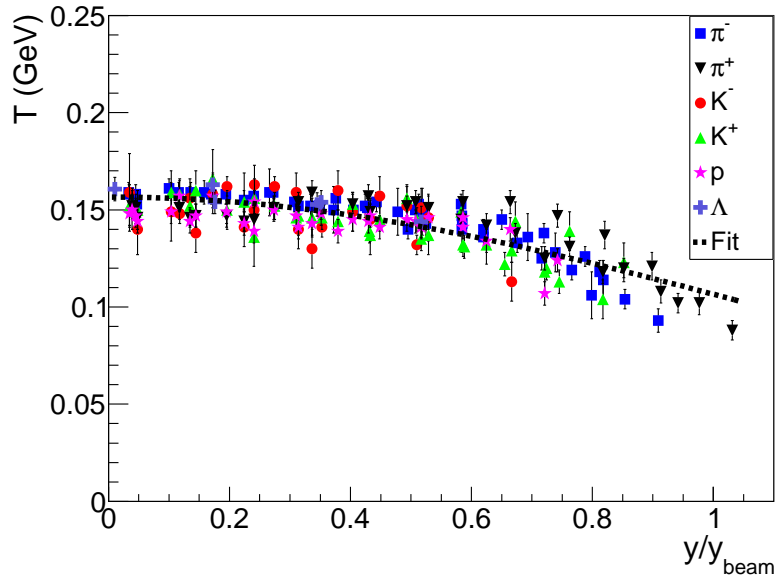}
}

\caption{The slope parameter $T$ for hadrons in bins of normalized rapidity $y/y_{beam}$
 from inelastic $pp$ collisions at 6.2-17.3 GeV. The experimental data are taken
 from~\cite{na61_hadrons},\cite{na61_hminus}, and~\cite{na61_lambda}.
 The dashed line indicate a fit to a Gaussian function (see text for details).
}
 \label{slopes}
\end{figure}

Transverse momentum distributions of hadrons from high-energy particle collisions are usually
described  by a Bose-Einstein distribution with the inverse slope parameter $T$ (effective temperature).
A collection of slope parameters values for hadrons from $pp$ collisions are plotted in
Fig.~\ref{slopes} as a function of normalized rapidity $y/y_{beam}$. The experimental data are taken
from~\cite{na61_hadrons},\cite{na61_hminus}, and~\cite{na61_lambda}, and for each hadron specie
we combined the measurements at all collision energies together. The rapidity dependence for the
slope parameter follows a Gaussian shape at all energies and for all hadrons.
It is interesting to note that for
each hadron we found little variation of the shape of the rapidity distributions for the slope
parameter with collision energy in the energy range from 6 to 17 GeV. Moreover, the shapes are
similar for all hadrons. The latter was checked by comparing the results of fitting of
the rapidity distribution for each hadron to a Gaussian: we found that the results for the fit
parameters vary by less than 3\% for the amplitude (the slope at the mean rapidity) and by
2\% for the standard deviation. A Gaussian shape with the averaged over all hadron specie parameters
($T(0)=$157~MeV, $\sigma=$1.1) is plotted in Fig.~\ref{slopes} by a dashed line,
and, indeed, as one can see, the measurements are tightly concentrated around this line. 

Antiprotons (not shown), however, do not follow the common trend.
The slope parameter for $\bar{p}$ appears to be rising (by about 40\%) in the NICA energy range,
still lying below  the common trend for other hadrons
(from 20\% to 60\% down depending on collision energy).

\section{Phase space distributions of hadrons from $pp$ reactions: data versus model}

\hspace{4mm}A number of feasibility studies to ensure optimal detector performance under
different experimental conditions at the NICA collider requires efficient and realistic
hadron phase space Monte Carlo generation. For this, dedicated software libraries - event
generators, may be used to predict high-energy particle physics events.
A list of microscopic models on the market includes UrQMD~\cite{urqmd}, EPOS~\cite{epos},
PHSD~\cite{phsd1}, and PHQMD~\cite{phqmd} generators. 
Despite of the fact that some models reproduce moderately well the energy dependence
of hadron yields, none of the existing event generators can simultaneously describe
total multiplicities, rapidity spectra, and transverse momentum distributions.   
In order to overcome this challenge we suggest a simple particle generator to obtain realistic
hadron phase space distributions using the results of the analysis of hadron production,
which were discussed in previous Sections.
The generator provides $\left<n\right>$ particles of a particular sort per event followed
a predefined profile along rapidity (Gaussian for produced mesons or polynomial for $p,\Lambda$)
and thermally distributed along transverse momentum. The parameters $\left<n\right>$,
$\sigma_{Gauss}$, and $T$ (slope) are taken from parameterizations discussed
in Sections~\ref{sec2},\ref{sec4}.
Owning to the fact that we are only interested in the phase space regions covered by the experimental
setup at the NICA collider, a short description of the detector and its simulation process
will be given.

\begin{figure}[hbt]
  \centerline{
    \includegraphics[width=85mm,height=58mm,angle=0]{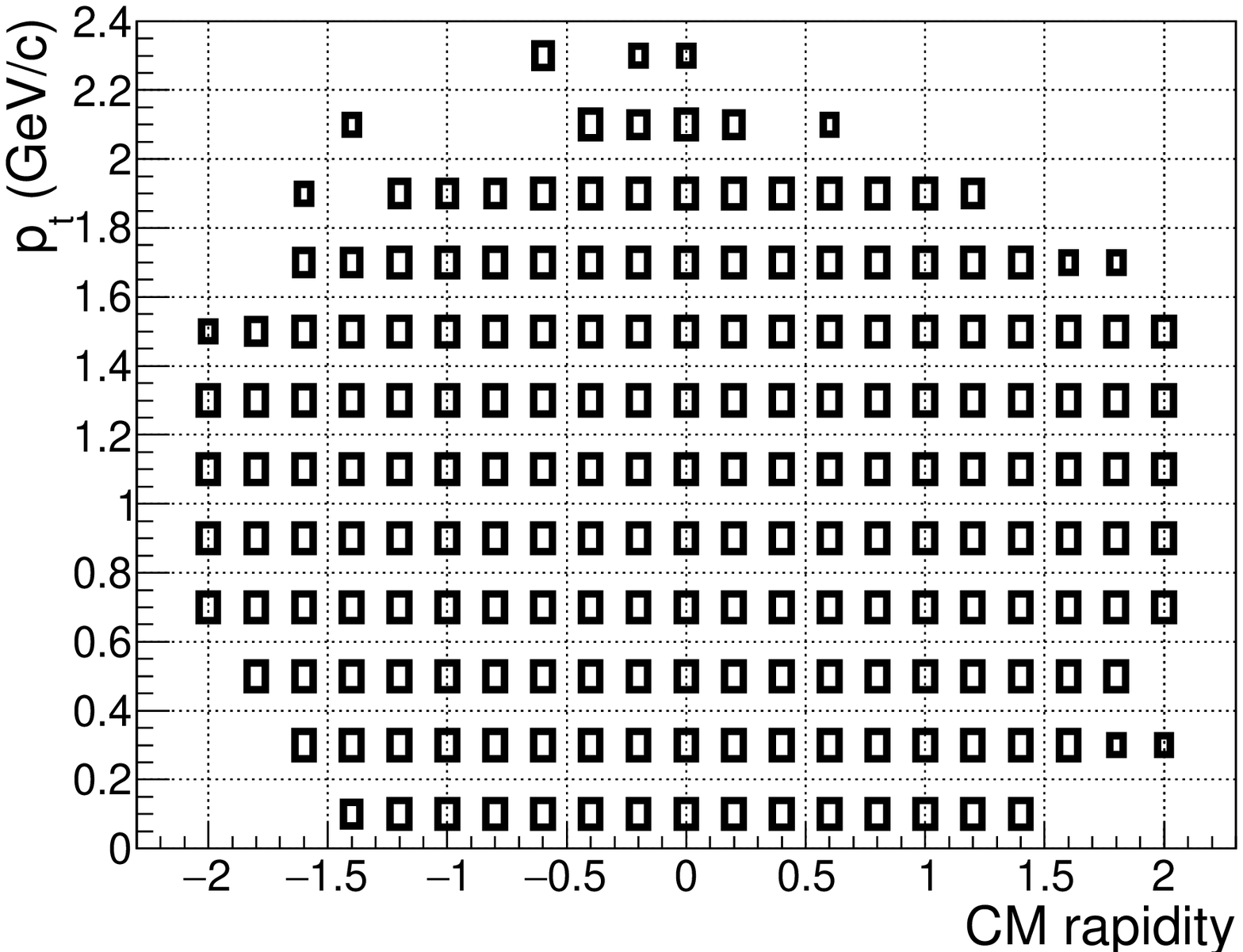}
    \includegraphics[width=85mm,height=60mm,angle=0]{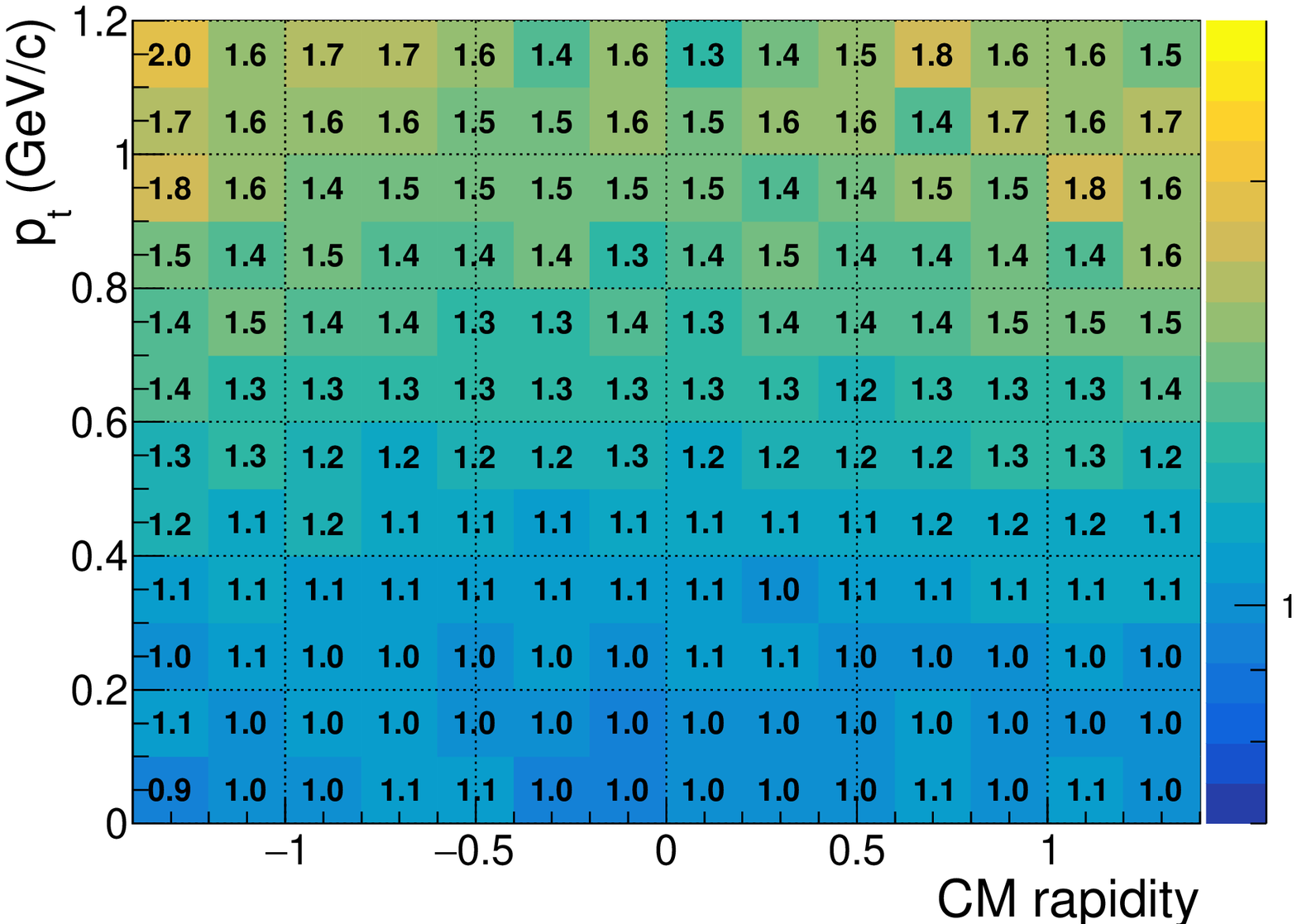}
}

\caption{Left: Phase space coverage of the MPD detector for positively charged kaons in $pp$
 collisions at $\sqrt{s}=9$~GeV in terms of center-of-mass rapidity and transverse momentum.
 The parameterization of the kaon phase space used as the simulation input is based on experimental
 data (see text for details). Right: The ratio of $K^+$ production rates $d^2N/(dp_tdy)$ from
 the PHQMD model~\cite{phqmd} and from the particle generator ($pp$ collisions at $\sqrt{s}=9$~GeV).    
 }
 \label{phasespace}
\end{figure}

The MPD (Multi-Purpose Detector) detector at NICA comprises large coverage of the reaction
phase space together with precise tracking and particle identification capabilities~\cite{nica_physics}.
Si\-mu\-la\-tion of high-energy particle interactions is performed within a dedicated software
framework - MPDRoot~\cite{mpdroot} which consists of a geometry toolkit for detector description,
track propagation routines based on the GEANT package, and multiple algorithms for detector response
simulation and event reconstruction. A set of dedicated interfaces, written in the implementation
language of ROOT (C++), allows easy transformation of the particle parameters from multiple
event generators to the input of the GEANT package to be propagated through the detector.
All the tracks which are registered in the detector elements are kept for further analysis.
The results of the simulation procedure for positively charged kaons with the suggested
particle phase space generator are shown in Fig.~\ref{phasespace} (left panel), where the phase space
coverage of kaons registered in the MPD setup is plotted in terms of center-of-mass rapidity
and transverse momentum. As one can see, the coverage in both longitudinal and transverse 
momentum components is sufficiently large, so reliable representation of hadron phase space
used as an input for simulation is, indeed, crucial.
The latter can be illustrated by comparing the predictions of the suggested particle generator
and different models.
In this study we compare the predictions of the suggested particle generator with
the PHQMD event generator.

The novel microscopic n-body dynamical transport approach PHQMD (Parton-Hadron-
Quantum-Molecular-Dynamics) is based on quark, diquark, string, and hadronic degrees of freedom
and extends the collision integral from the established PHSD (Parton-Hadron-String-Dynamics)
approach~\cite{phsd1, phsd2} uniting it with 2-body potential interactions between baryons.
While high energy inelastic hadron-hadron collisions in PHQMD are described by the FRITIOF string
model~\cite{fritiof}, the transport approach at low energy hadron-hadron collisions is based
on experimental cross sections and matched to reproduce the nucleon-nucleon, meson-nucleon
and meson-meson cross section data in a wide kinematic range. In the string fragmentation process,
implemented in the model, strangeness production is governed by addition parameters, which define
the probability to create a strange quark-antiquark pair or diquark-antidiquark pair. 
In Fig.~\ref{phasespace} (right panel) is shown the ratio of $K^+$ production rates ($d^2N/(dp_tdy)$)
from the PHQMD model and from our particle phase space generator.
As one can see, the model reasonable well
describes the kaon production near midrapidity and at low transverse momenta. At larger momenta,
however, the difference in the production rates between experimental data and model is growing
progressively, thus, overall transverse dynamics in the PHQMD generator needs some additional tuning
to reproduce data.

Finally, we can point out that the suggested simple particle phase space generator, which
describes hadron production phase space in $pp$ collisions based on experimental data,
can be very useful for testing available event generators and performing detector performance
studies in the NICA energy range.   
 
\section{Summary}

\hspace{4mm}Using existing experimental data for elementary $pp$ collisions,
a new evaluation of the energy dependence of hadron production
is performed within the NICA energy range. Results for mean multiplicities,
rapidity spectra, and transverse momentum distributions of $\pi^{\pm}$,
$K^{\pm}$, $K^{0}_S$, $\Lambda$, $p$, $\bar{p}$ are collected in the region of collision
energies $3<\sqrt{s_{NN}}<31$~GeV.
This new collection includes recent measurements from the CERN NA49 and NA61 experiments
for multiple hadron species. The excitation function of the particle yields is analyzed against
Redge- and Lund-model motivated fits and new parameterizations for hadron production cross-sections
in inelastic proton-proton interactions are obtained. These results can be used for testing and tuning
 available microscopic models.  
In the study of the variation of the rapidity and transverse momentum spectra of hadrons with energy,
an interesting scaling behavior of relevant parameters at NICA energies is observed. Based on this
observation, a standalone algorithm for hadron phase space generation in $pp$ collisions is suggested.
This particle phase space generator was integrated into the MPD detector simulation framework and compared
with the novel PHQMD generator. 

\section{Acknowledgments}
\hspace{4mm}This work was supported by the Russian Scientific Fund Grant 19-42-04101.
Furthermore, we acknowledge support by the Deutsche Forschungsgemeinschaft (DFG, German Research Foundation).

\end{document}